\documentstyle[12pt]{article}
\def\aprime {\alpha^{\prime}}
\def\half {{1 \over 2}}
\def\Tr {\mbox{Tr}}

\def\np    { Nucl. Phys. }
\def\pr    { Phys. Rev. }
\def\pl    { Phys. Lett. }
\def\cmp   { Commun. Math. Phys. }

\def\ijmp  { Int. J. Mod. Phys. }

\makeatletter
\@addtoreset{equation}{section}

\def\del {\partial}

\def\calH {{\cal H}}

\def\be{\begin{equation}}     
\def\ee{\end{equation}}
\def\bea{\begin{eqnarray}}     
\def\eea{\end{eqnarray}}
\def\&{&\!\!\!\!\!\!\!\! &}

\def\nn{\nonumber}

\def\parmedskipn        {  \par\medskip\noindent  }

\renewenvironment{thebibliography}{\pagebreak[3]\par\vspace{0.6em}
\begin{flushleft}{\large \bf References}\end{flushleft}
\vspace{-1.0em}

\begin{enumerate}\if@twocolumn\baselineskip=0.6em\itemsep -0.2em
\else\itemsep -0.2em\fi\labelsep 0.1em}{\end{enumerate}}

\begin{document}
\baselineskip=0.65cm



\begin{titlepage}

    \begin{normalsize}
     \begin{flushright}
                 NSF-ITP-01-14\\
                 SU-ITP-01/07\\
                 hep-th/0102174\\
                 February 2001
     \end{flushright}
    \end{normalsize}
    \begin{LARGE}
       \vspace{1cm}
       \begin{center}
        Metamorphosis Of Tachyon Profile\\
        In Unstable $D9$-Branes
       \end{center}
    \end{LARGE}
  \vspace{5mm}

\begin{center}
           Koji Hashimoto
           \footnote{E-mail address:
              koji@itp.ucsb.edu}
            and Shinji Hirano
           \footnote{E-mail address:
              hirano@itp.stanford.edu}   \\
      \vspace{4mm}
        ${}^*${\it Institute for Theoretical Physics,} \\
        {\it University of California, Santa Barbara,
             CA 93106}\\
        ${}^\dag${\it Department of Physics, Stanford University,}\\
        {\it Stanford, CA 94305}\\
      \vspace{1cm}

     ABSTRACT 
        \par
\end{center} 
\begin{quote}
 \begin{normalsize}
We explored a variety of brane configurations in our previous paper
within the two derivative truncation of the unstable $D9$-brane
effective theory. In this paper we extend our previous results with
emphasis on the inclusion of the higher derivative corrections for the
tachyon and the gauge fields computed in the boundary string field
theories. We give the exact solutions to BPS brane configurations
studied in our previous paper and find remarkable exact agreements of
their energies and RR-charges with the expected results. We further
find a few more solutions that we could not construct in the two
derivative  truncations, such as a ($F$,$D6$) bound state ending on a
$D8$-brane whose existence turns out to be due to a higher derivative
effect and also the dielectric brane of Emparan and Myers as a
nonsupersymmetric example. These are also in exact agreements with 
the results obtained in the effective theory of supersymmetric D-branes.

 \end{normalsize}
\end{quote}

\end{titlepage}
\vfil\eject

\setcounter{footnote}{0}




\section{Introduction}

The low energy effective theories of supersymmetric $D$-branes have
been providing us with a tremendous amount of developments in string
theory. The intuition based on the stingy pictures of $D$-branes
illuminated not only various nonperturbative phenomena in
supersymmetric gauge theories but also the statistical origin of the
black hole entropy. This line of progress is highlighted by the
duality between gravity and gauge theory such as the Matrix theory and
the $AdS/CFT$ duality. The idea of the near horizon limit in the
$AdS/CFT$ duality elucidated the precise correspondence of gauge
theories and their supergravity duals. Although these dualities
suggest conceivable conjectures for the nonperturbative definition of
string theories by gauge theories, they would define at most certain
corners of a vast moduli space of the whole string theories. So it
would be anticipated to search for new ideas which might make further
progress in the nonperturbative formulation of string theory.  

A potentially attractive idea is that the unstable $D9$-brane systems
would be so protean that they could reproduce all kinds of branes
in superstring theories dynamically as kinks or lumps and even they
could tell us about the closed string vacuum via the tachyon
condensation\cite{Sen1, Sen2, Sen3}. Actually some exact results that
can make this idea realized more concretely have been provided by the
boundary string field theories (BSFT)\cite{BSFT, Shata} in 
\cite{Gerasimov, KMMbos, KMM, andreev, Sigma, KL, TTU} and also by the 
noncommutative tachyons\cite{DMR, KHLM} in the wake of remarkable 
results\cite{SZ,BSZ} of the cubic open string field theory\cite{cubic} 
in the level truncation\cite{KSamu}. The BSFT gives the
exact form of the tree level tachyon potential, and the exact
solutions for the flat $D$-branes are found both in the
BSFT\cite{KMMbos, KMM} and the noncommutative tachyon
approach\cite{DMR, KHLM}. The flat $D$-branes, 
however, are somewhat too simple to 
invoke the protean nature of the unstable $D9$-branes. So in this
paper we will study nontrivial brane configurations explored in our
previous paper\cite{KS} in the two derivative truncation\footnote
{up to a term which depends on the renormalization 
scheme\cite{Sigma} }, by making use of the BSFT action with the 
inclusion of the higher derivative corrections for the tachyon and 
the gauge fields. Remarkably, even for nontrival brane configurations, 
we find exact agreements with the expected superstring results. 
Our results show nontrivial metamorphoses of the tachyon profile 
of more involved kinks or lumps than those previously considered in 
the BSFT literatures, thus giving a little step further to push the 
protean nature of unstable $D$-branes.  

As is quite different from the effective theory of supersymmetric
$D$-branes, we cannot ignore the higher derivative corrections in the
effective theory of unstable branes even in the low energy limit. When
the energy scale gets much smaller than the string scale, the tachyon
mass square becomes infinitely negative, indicating a violent
destablization of the system, that is, a potential with an infinitely
negative curvature. The tachyon field will be simply frozen at the
deep bottom of the potential. To keep the nontrivial dynamics of this
tachyonic system, the tachyon must be fluctuating wildely against the
free fall in the tachyon potential. This implies significance of the
higher derivative corrections for the tachyon in the low energy
effectively theory of unstable $D$-branes. We can see it explicitly in
the BSFT results. The solutions representing the flat $D$-branes have
the linear profile, whose coefficient must be taken to infinity to
give the minimum of the energy and to find the exact agreement with the
expected results. Thus the higher the derivative corrections are, the
more they give the dominant contributions in the low energy effective
theory of unstable $D$-branes in the BSFT. 

It seems somewhat awful and intractable to deal with the action that
includes infinite number of the higher derivative corrections. However
it is quite tractable, as we deduced several nontrivial exact results
from the higher derivative corrected action of the BSFT, which
suggests the usefullness of the effective theory of unstable
$D$-branes, though it may not be as powerful as the low energy
effective theory of supursymmetric $D$-branes. In particular we only
worked out the classical analysis, that was sufficient in this paper,
as we mostly focused on the BPS configurations which are believed not
to be subject to quantum corrections, though out of nonsupersymmetric
$D9$-branes. But in general we have to take into account the quantum
corrections. It is very much so in particular in such a theory like
the effective theory of unstable or nonsupersymmetric $D$-branes that
we are working on. It, however, obviously seems quite hard to carry
out the loop computations in the awfully higher derivative corrected
action. This is a big drawback of the tachyon models. 
 But we would like to emphasize that there are still so much things to
 be done even in the classical analysis of unstable $D$-branes. At
 least we are able to reproduce many other brane configurations found
 in the supersymmetric $D$-brane analysis, though we always have to
 keep in mind an important question whether the tachyon models could
 go beyond what we have already done. 
 
The organization of our paper is as follows. In section \ref{action},
we start with the BSFT action and list a
couple of special cases  of the action which will be useful in the
subsequent sections. Also we make general remarks on the equations of
motion for the tachyon and the gauge fields. In section \ref{warmup},
we give a few examples of simple generalizations of a kink solution,
including a $D6$-$D8$ and ($F$,$D8$) bound state. Also we discuss a
trivial example of a description of the fundamental strings at the
closed string vacuum. In section \ref{beb}, we turn to nontrivial
examples such as $D6$-branes ending on a $D8$-brane and its
generalization to a ($F$,$D6$) bound state ending on a $D8$-brane that 
we could not construct in our previous paper\cite{KS}. In
section \ref{turningonB}, we provide a noncommutative generalization
of the configuration discussed in section \ref{beb}. In section
\ref{junction}, we return to a configuration akin to the type
discussed in section \ref{warmup}, that is, a junction
configuration. In section \ref{emparanmyers} we study the
Emparan-Myers' effect\cite{Emparan, Myers} as a nonsupersymmetric
example which we did not 
discuss in our previous paper\cite{KS} either. In section \ref{nonabelian}, 
we argue a nonabelian generalization of the configuration discussed in 
section \ref{beb}, that is, a $D6$-brane suspended between two
$D8$-branes. Finally we will close our paper with summary and
discussions in section \ref{discussion}. 

\section{The action}
\label{action}

We will employ the results of BSFT for superstring
theories\cite{KMM,andreev,Sigma,KL,TTU}. When the tachyon $T$ takes the form
of the linear profile $T=qx$ and the field strength of the gauge
fields are constant, there is a complete result including the
couplings of the gauge fields and the derivatives of the
tachyon\cite{andreev}. So the action we are going to use will be valid
up to the second derivatives such as $\del^2T$ and $\del F$. 
In this paper, however, we will assume that we could apply the BSFT 
action even for nontrivial configurations of the tachyon $T$ and the
gauge fields $A_{\mu}$, though it would certainly go beyond the  
validity of the BSFT results. But we will mostly consider BPS
configurations, so it is conceivable for the BSFT action to be still
valid due to the nonrenormalization theorems of supersymmetric
configurations. Now the BSFT action is given by
\begin{eqnarray}
S&=&-T_{D9}\int dtd^9x\Tr\Biggl(e^{-2\pi\aprime T^2}
\det\sqrt{2\pi}\nn\\
&&\times\left.
\frac{\prod_{r=1/2}^{\infty}\det
\left(\eta_{\mu\nu}+2\pi\aprime F_{\mu\nu}
+4\pi(\aprime)^2{D_{(\mu}TD_{\nu)}T \over r}\right)}
{\prod_{n=1}^{\infty}\det
\left(n\left(\eta_{\mu\nu}+2\pi\aprime F_{\mu\nu}\right)
+4\pi(\aprime)^2D_{(\mu}TD_{\nu)}T\right)}\right),
\label{theaction}
\end{eqnarray}
where the $\zeta$-function regularization is implied. The quadratic
term $D_{(\mu}TD_{\nu)}T$ of the covariant derivatives of the tachyon
is symmetrized with respect to their spacetime indices, so that our
generalized action will be consistent with the result in \cite{KMM}
for the tachyon configuration $T=q\Gamma^ix_i$ of the higher
codimension $D$-branes. There is a caveat concerning the ordering of
$U(N)$ matrices inside the trace in the case of $N$ multiple
$D9$-branes, which is the same problem as the one in the nonabelian
Dirac-Born-Infeld (DBI) action\cite{nonAbelian}. For the most part of
our paper, we will only consider a single $D9$-brane, so this problem
will not be in our concern. But we will discuss two $D9$-brane system
only in section \ref{nonabelian}, where we will make an ansatz as for
this ordering.  

Now we will list the forms of the action (\ref{theaction}) in a few
special cases (only abelian cases) that we are going to consider in
the most part of our paper. When the gauge fields are vanishing, the
above action is simplified to  
\begin{eqnarray}
S=-T_{D9}\int dtd^9x e^{-2\pi\aprime T^2}
F[4\pi(\aprime)^2\del_{\mu}T\del^{\mu}T],
\label{simple}
\end{eqnarray}
where $F[x]=\frac{x4^x\left(\Gamma(x)\right)^2}{2\Gamma(2x)}$ 
\cite{KMM}.

Next let us consider nonvanishig gauge fields. When we turn on the
field strengh $F_{\mu\nu}$ only in the directions orthogonal to those
directions on which the tachyon depends nontrivially, the action
factorizes to\cite{AFTT}
\begin{eqnarray}
S=-T_{D9}\int dtd^9x e^{-2\pi\aprime
  T^2}F[4\pi(\aprime)^2\del_{\mu}T\del^{\mu}T]
  \sqrt{-\mbox{det}\left(\eta_{\mu\nu}+ 
  2\pi\aprime F_{\mu\nu}\right)}.
\label{factorize}
\end{eqnarray}
In a more specific case that the gauge fields in 3-dimensional space,
labeled by the coordinates $\{y_6,y_7,y_8\}$, are turned on 
and the tachyon depends only on the coordinates $\{y_6,y_7,y_8,x_9\}$, 
we have a little more involved action  
\begin{eqnarray}
S&=&-T_{D9}\int dtd^9xe^{-2\pi\aprime T^2}
\sqrt{-
\det\left(\eta_{ab}+2\pi\aprime F_{ab}\right)}\nn\\
&&\times
F\left[4\pi(\aprime)^2\!\left\{\!(\del_9T)^2 \!+\!
\frac{\left((\del_iT)^2
\!-\!(2\pi\aprime\epsilon_{ijk}E_j\del_kT)^2\right)
\!+\!(2\pi\aprime B_i\del_iT)^2}
{-\det\left(\eta_{ab}+2\pi\aprime F_{ab}\right)}\!
\right\}\right],
\label{involved}
\end{eqnarray}
where $a,b=0,6,7,8$, whereas $i,j=6,7,8$, and 
$-\det\left(\eta_{ab}+2\pi\aprime F_{ab}\right)=1+\half 
(2\pi\aprime F_{ab})^2-(2\pi\aprime)^4(E_iB_i)^2$.

\parmedskipn
\underline{The equations of motion}

\noindent
In the following sections we will give a variety of solutions of the
equations of motion (EOM) derived from the above action
(\ref{theaction}). In all the solutions we will explore below, the
tachyon $T$ is typically proportional to a constant $q$ which is taken
to $\infty$. It means physically that the brane configurations
described by these solutions are localized in space in certain ways,
due to the tachyon potential $e^{-2\pi\aprime T^2}$, an overall factor
in the action. Now because of this specific property of the tachyon
$T$, the EOM for the tachyon is simplified quite a bit. When $q$ goes
to $\infty$, we can ignore the contributions of lower orders in
$q$. Then the tachyon EOM boils down to 
\begin{eqnarray}
F - \del_{\mu}T\frac{\delta F}{\delta\del_{\mu}T}=0,
\label{tachyonEOM}
\end{eqnarray}
where $F$ is the functional in particular of $(\del T)^2$ appearing in
eqs. (\ref{simple}),(\ref{factorize}) and (\ref{involved}). As we will
see, when the tachyon EOM (\ref{tachyonEOM}) is satisfied, it is
tantamount to the identity 
\begin{equation}
F[a]=2aF'[a]\qquad \mbox{for}\qquad a\to\infty.
\label{identity}
\end{equation}

As for the gauge fields, we only turn on the electric and magnetic
fields, $E_i$ and $B_i$ ($i=6,7,8$), in (3+1)-dimensional subspace of
the (9+1)-dimensional spacetime. Thus for later applications the gauge
EOMs take the forms 
\def\calL {{\cal L}}
\begin{eqnarray}
\epsilon_{ijk}\del_j\frac{\delta\calL}{\delta B_k}&=&0,
\label{magneticEOM}\\
\del_i\frac{\delta\calL}{\delta E_i}&=&0,
\label{electricEOM}
\end{eqnarray}
where $\calL$ is the Lagrangian density of the above action.

\parmedskipn
\underline{The Chern-Simons couplings}

\noindent
We will also compute the RR-charges of our brane configurations. So we
give a formula for the Chern-Simons (CS) coupling on the unstable
$D9$-branes computed in \cite{KL,TTU}: 
\begin{eqnarray}
S_{\rm CS}=T_{D9}\int C\wedge\Tr e^{2\pi\aprime (F-T^2+DT)}.
\end{eqnarray}
%

\section{The warm-ups}
\label{warmup}
We first discuss a couple of simple generalizations of a kink solution
$T={q \over \sqrt{\aprime}}x$ with $q\to\infty$ which describes a
single $D8$-brane sharply localized at $x=0$ in the 9th-direction. 

\parmedskipn
\underline{$D6$-$D8$ bound state}\footnote{The $D$-brane bound state 
of this type was previously discussed in \cite{AFTT}.}

\noindent
The $D6$-branes parallel to $D8$-branes are unstable against
$D6$-branes dissolving into $D8$-branes, and thus they are melting
into $D8$-branes uniformly. From the $D8$-brane viewpoint, this
non-marginal bound state can be described by a uniform magnetic field
on the $D8$-branes, as is obvious from the CS couplings of
the RR-fields. Let us turn on a constant magnetic field $B_6$, which
represents $D6$-branes uniformly distributed over the (7,8)-plane on a
$D8$-brane. It can be easily checked that 
\begin{eqnarray}
T&=&{q \over \sqrt{\aprime}}x,\qquad q\to\infty\\
B_6&=&\mbox{const.},
\end{eqnarray}
is a solution of the EOM derived from the action
(\ref{factorize}). The energy of the bound state is readily evaluated
as 
\begin{eqnarray}
E=\sqrt{1+(2\pi\aprime B_6)^2}\;T_{D8}\int d^8y,
\label{ed6d8}
\end{eqnarray}
where $T_{D8}=T_{D9}\sqrt{2\pi^2\aprime}$ and we used the asymptotics
of $F[4\pi\aprime q^2]$ at large $q$, 
\def\calO {{\cal O}}
\begin{equation}
F[4\pi\aprime q^2]=\sqrt{4\pi^2\aprime}q
+{1 \over 8}\frac{\sqrt{\pi}}{\sqrt{4\pi\aprime}q}+\calO(q^{-3}).
\label{asymptotics}
\end{equation}
Also the $D6$-brane charge is computed as
\begin{eqnarray}
&&T_{D9}\int e^{-2\pi\aprime T^2}(2\pi\aprime)^2dT\wedge F \nn\\
&=&T_{D6}(B_6/2\pi)\int dy_7dy_8,
\end{eqnarray}
where $T_{D6}=T_{D9}2(2\pi^2\aprime)^{3/2}$, and $B_6/2\pi$ gives the
number of $D6$-branes $N_{D6}$. Note that the energy (\ref{ed6d8}) of 
the $D6$-$D8$ bound state has the correct value, as we can see that if 
the $D8$-brane were absent in the energy formula (\ref{ed6d8}), we
would correctly get the tension $T_{D9}2(2\pi^2\aprime)^{3/2}N_{D6}$
of $N_{D6}$ $D6$-branes.  

\vspace{5mm}

\def\calF {{\cal F}}
\parmedskipn
\underline{($F$,$D8$) bound state}

\noindent
Similarly the fundamental string ($F1$) on $Dp$-branes can be
described by an electric flux on them, as can be seem from the
coupling of NS-NS 2-form $B^{\rm NS}$ with the invariant field
strength $\calF=2\pi\aprime (B^{\rm NS}+F)$. Let us trun on a constant
electric field $E_6$, which represents the fundamental strings lying
on the (0,6) plane. Again it is easy to check that  
\begin{eqnarray}
T&=&{q \over \sqrt{\aprime}}x\qquad q\to\infty,\\
E_6&=&\mbox{const.},
\end{eqnarray}
is a solution of the EOM derived from the action (\ref{factorize}).

Now let us compute the energy. The conjugate momentum $\Pi_A$ of the
gauge field $A_6$ is given by 
\begin{eqnarray}
\Pi_A=\frac{(2\pi\aprime)^2E_6}
{\sqrt{1-(2\pi\aprime E_6)^2}}T_{D8}V_8,
\end{eqnarray}
where $V_8$ is the volume of a $D8$-brane. Note that the gauge fields
are $U(1)$-valued and thus compact, so the conjugate momentum $\Pi_A$
is quantized as $Nl$ where $l$ is the length of the fundamental
strings. Then the Hamiltonian $H$ is computed as
\begin{eqnarray}
H&=&\Pi_A E_6 - L\\
&=&\sqrt{(T_{D8}V_8)^2+\left(Nl/2\pi\aprime\right)^2}.
\end{eqnarray}
This gives the exact agreement.

\vspace{5mm}

\parmedskipn
\underline{A trivial example of the fundamental strings}

\noindent
It is one of the important issues in the tachyon models how to
describe the fundamental strings at the closed string
vacuum. Following Yi\cite{Yi}, there are several attempts to argue
that the fundamental strings come about as confined fluxes on the
unstable branes\cite{BHY,GHY,KLS}. Here we will give a trivial description
of the fundamental strings, simply by turning on a constant electric
field at the closed string vacuum. Thus they are given by 
\begin{eqnarray}
T&=&\pm\infty,\\
E_9&=&\mbox{const.}.
\end{eqnarray}
This is obviously a solution of the EOMs. Now let us compute the
energy of this configuration. The Lagrangian in this case is 
\begin{eqnarray}
L=T_{D9}V_9e^{-2\pi\aprime T^2}
\sqrt{1-(2\pi\aprime E_9)^2},
\end{eqnarray}
where $V_9$ denotes the volume of the 9-dimensional space.
So the Lagrangian is vanishing for this configuration. However the
Hamiltonian has a finte value due to the quantization condition of the
conjugate momentum $\Pi$ of the gauge field $A_9$, which is given by 
\begin{eqnarray}
\Pi={\del L \over \del E_9}
=T_{D9}V_9e^{-2\pi\aprime T^2}
\frac{(2\pi\aprime)^2E_9}{\sqrt{1-(2\pi\aprime E_9)^2}}
=Nl,
\end{eqnarray}
where $l$ is the length of the fundamental strings. Thus the
Hamiltonian is computed as 
\begin{eqnarray}
H&=&\Pi E_9-L
=\sqrt{\left(T_{D9}V_9e^{-2\pi\aprime T^2}\right)^2
+\left({Nl \over 2\pi\aprime}\right)^2}\nn\\
&=&{Nl \over 2\pi\aprime},
\end{eqnarray}
that is exactly the energy of the $N$ fundamental strings.


\section{Branes ending on branes}
\label{beb}

We move on to nontrivial examples where the gauge fields are not as
simple as those discussed in the previous section, but they take
nontrivial configurations such as monopoles and dyons. It is quite
helpful to first consider the problem within a simple approximation
and to guess the form of the solutions in the fullfledged
treatment. It often happens for the BPS configurations that the
linearized approximation will give the same result as even when the
nonlinear couplings are included, as is actually the case for the BPS
BIon of \cite{CM, Gary}. The configuration we consider
in this section is exactly of the type of BPS BIons, that is,
$D6$-branes ending on a $D8$-brane and a ($F$,$D6$) bound state ending
on a $D8$-brane. So we will first do the linearized approximation. The
recipe for obtaining the solution goes as,  
(1) starting with a $D8$-brane solution, given by a kink on an
unstable $D9$-brane,  
(2) computing the collective excitations about a kink,
(3) assuming the massive collective modes will not contribute to the
configuration we are looking for, 
(4) doing the linearized approximation for the zero modes,
(5) solving the EOM for the zero modes 
and (6) checking if the solution obtained in the linearized
approximation really satisfies the EOM of the fullfledged action. As
we will see, happily it works. 

\subsection{$D6$-branes ending on a $D8$-brane}
\label{d6endingd8}

\parmedskipn
\underline{The linearized approximation}

\def\Ttilde {\tilde{T}}
\noindent
Now we are going to look at the collective excitations about a kink
representing a $D8$-brane. The results are given by Minahan and
Zweibach\cite{Zwie, MZ, MZ2} in eq.(6.8) and (2.29) in their 
paper \cite{MZ2}. For the massless modes of the tachyon, denoting
$T=\frac{q}{\sqrt{\aprime}}x+\Ttilde$ (here $\Ttilde$ is restricted to
the zero modes for our purpose, but in general it denotes all the
collective excitations), we have the linearlized action, 
\begin{eqnarray}
S_T=-T_{D9}\int dtd^8ydxe^{-2\pi q^2x^2}\left[F[4\pi\aprime q^2]
+ 4\pi(\aprime)^2\frac{F[4\pi\aprime q^2]}
{4\pi\aprime q^2}\half(\del_{\mu}\Ttilde)^2\right].
\end{eqnarray}
For the massless modes of the gauge field collective excitations, 
we have
\begin{eqnarray}
S_A=-T_{D9}\int dtd^8ydxe^{-2\pi q^2x^2}
F[4\pi\aprime q^2]{1 \over 4}
(2\pi\aprime)^2F_{\mu\nu}F^{\mu\nu},
\end{eqnarray}
where the indices $\mu,\nu$ denote the directions transverse to a
$D8$-brane or $y$-directions. 
Therefore 
in the linearized approximation our problem reduces to that of BIons
\cite{CM, Gary}, and we can readily read off the solution of the EOM for
zero modes. The result for large $q$ is  
\begin{eqnarray}
\Ttilde&=&2\pi qNc_m{\sqrt{\aprime} \over r},\\
B_i&=&Nc_m{y_i \over r^3},
\end{eqnarray}
where $B_i (i=6,7,8)$ denotes a magnetic field on a $D8$-brane, and
$c_m$ and $N$ is a numerical constant and the number of $D6$-branes
respectively. Note that we chose the power of $q$ in the front
coefficients from the normalizations of the tachyon and gauge kinetic
terms. 

In sum, our conjecture for the configuration, $N$ $D6$-branes ending
on a $D8$-brane, is given by 
\begin{eqnarray}
T&=&\frac{q}{\sqrt{\aprime}}\left(x+2\pi Nc_m{\aprime \over r}\right),
\label{tachyon}\\
B_i&=&Nc_m{y_i \over r^3}\quad (F=Nc_md\Omega_2),
\label{gauge}\\
&&\mbox{with}\qquad q\to\infty \nn
\end{eqnarray}
as is proposed in \cite{KS} in the two derivative truncation.

\parmedskipn
\underline{The energy}

\noindent
Now let us determine the value of a numerical constant $c_m$. To
reproduce the energy of $D6$-brane correctly, it turns out that
$c_m=\half$, in the linearlized approximation. The energy is given by
\begin{eqnarray}
E&=&T_{D9}\int d^8ydx e^{-2\pi q^2x^2}
F[4\pi(\aprime)^2 q^2]
\left[1+(2\pi\aprime)^2(Nc_m)^2{1 \over r^4}\right]\nn\\
&=&T_{D9}\sqrt{2\pi^2\aprime}\int d^8y
+T_{D9}\sqrt{2\pi^2\aprime}\Omega_2(2\pi\aprime)^2(Nc_m)^2
\int d^5y\int_{0}^{\infty}{dr \over r^2}.
\end{eqnarray}
Noting that $\Omega_2=4\pi$ and $x=-2\pi\aprime Nc_m/r$, we find that,
when $c_m=\half$, we exactly reproduce the sum of the energies of a
$D8$ and $N$ $D6$-branes: 
\begin{eqnarray}
E=T_{D9}\sqrt{2\pi^2\aprime}\int d^8y
+T_{D9}2(2\pi^2\aprime)^{3/2}N\int d^5y\int_{-\infty}^0dx.
\end{eqnarray}
Remark also that $D6$-branes are extended only on the half line in
$x$-direction, as it should be, being emphasized in \cite{KS}. 

\parmedskipn
\underline{The RR-charge}

\noindent
Let us compute the RR-charges of our configuration. Recalling the CS
coupling  
\begin{eqnarray}
S_{\rm CS}=T_{D9}\int C\wedge\Tr e^{2\pi\aprime (F-T^2+DT)}
\label{CS}
\end{eqnarray}
on the unstable $D9$-branes, 

\noindent
(1) the $D8$-brane charge is given by
\begin{eqnarray}
T_{D9}\int C^{(9)}\wedge e^{-2\pi\aprime T^2}(2\pi\aprime)dT
&=&T_{D9}q\int C^{(9)}\wedge e^{-2\pi q^2(x+{\pi\aprime N \over r})^2}
(2\pi\sqrt{\aprime})d(x+{\pi\aprime N \over r})\nn\\
&=&\sqrt{2}\pi\sqrt{\aprime}T_{D9}\int C^{(9)}\wedge
\delta(x+{\pi\aprime N \over r})
d(x+{\pi\aprime N \over r})\nn\\
&=&\sqrt{2}\pi\sqrt{\aprime}T_{D9}\int C^{(9)}.
\end{eqnarray}
(2) The $D6$-brane charge is similarly evaluated as\footnote
{We assumed a spherical symmetry of the 7-form 
$C^{(7)}$.}
\begin{eqnarray}
T_{D9}\int C^{(7)}\wedge e^{-2\pi\aprime T^2}(2\pi\aprime)^2
dT\wedge F =T_{D9}2(2\pi^2\aprime)^{3/2}N\int C^{(7)}.
\end{eqnarray}
We find the exact agreement with the RR-charges.

\parmedskipn
\underline{The fullfledged treatment: nonlinear}

\noindent
Having found the exact agreement of the energies and the RR-charges for
the solution in the linearized approximation\footnote{The Chern-Simons 
coupling (\ref{CS}) given in \cite{KL,TTU} is exact in the full
superstring theory.}, it is natural to suspect 
that it will be the exact solution of the whole nonlinear action
(\ref{involved}), just like in the case of BPS BIons\cite{CM,
  Gary}. We are going to show that it is indeed the case.  

\noindent
(1) The tachyon EOM:

\noindent
As noted before, the tachyon EOM is simplified to eq.\
(\ref{tachyonEOM}). Now the argument of the functional $F$ in this
case is given by 
\begin{eqnarray}
4\pi(\alpha')^2
\left(
(\del_9T)^2 +
\frac{(\del_iT)^2+(2\pi\aprime B_i\del_iT)^2}
{1+(2\pi\aprime B_i)^2}
\right)
\!\!\!\!&&=
4\pi(\alpha')^2
\left(
(\del_9T)^2 +(\del_iT)^2
\right)\nonumber \\
&&=4\pi\aprime q^2
\left(1+\frac{(\pi N\aprime)^2}{r^4}\right),
\end{eqnarray}
where the first equality is due to the Bogomol'nyi-like equation
$\del_iT=-{q \over \sqrt{\aprime}}2\pi\aprime B_i$, which was indeed
the Bogomol'nyi equation in the two derivative
truncation\cite{KS}. The tachyon EOM further reduces to the identity
(\ref{identity}), 
\begin{eqnarray}
&&F[a]=2aF'[a],\\
&&\mbox{with}\qquad a=4\pi\aprime q^2
\left(1+\frac{(\pi N\aprime)^2}{r^4}\right),
\end{eqnarray}
that indeed holds when $a\to\infty$ or equivalently $q\to\infty$.

\noindent
(2) The gauge field EOM:

\noindent
Again as we remarked before, the gauge EOM in this case is simply the
one (\ref{magneticEOM}) for the magnetic field, 
\begin{eqnarray}
\epsilon_{ijk}\del_j\frac{\delta\calL}{\delta B_k}=0.
\label{gaugeeom}
\end{eqnarray}
It can readily be seen that this EOM holds, as the l.h.s. is
proportional to $y_jy_k-y_ky_j$, which is identically zero. 

\parmedskipn
\underline{The energy: in the fullfledged treatment}

\noindent
Now we will show the energy evaluated from the fullfledged nonlinear
action (\ref{involved}) is exactly the same as the linearized
one. Again noting the asymptotics (\ref{asymptotics}) of the
functional $F$, the energy (\ref{involved}) takes the form
\begin{eqnarray}
E\!\!\!&=&\!\!\!T_{D9}\int d^8ydx
e^{-2\pi q^2(x+{\pi\aprime N \over r})^2}
\sqrt{4\pi^2\aprime}
q\sqrt{1+\frac{(\pi N\aprime)^2}{r^4}}
\times\sqrt{1+\frac{(\pi N\aprime)^2}{r^4}}.
\end{eqnarray}
Remarkably the square root becomes the perfect square, as it should be
in the case of BPS configurations. Let us finish up the computation: 
\begin{eqnarray}
E&=&T_{D9}\sqrt{2}\pi\sqrt{\aprime}\int d^8y
+T_{D9}\sqrt{2\pi^2\aprime}4\pi(\pi N\aprime)^2
\int d^5ydx{dr \over r^2}\delta(x+{\pi N\aprime \over r})\nn\\
&=&T_{D9}\sqrt{2}\pi\sqrt{\aprime}\int d^8y
+T_{D9}2(2\pi^2\aprime)^{3/2}N
\int d^5ydx\theta(-x).
\end{eqnarray}
This is exactly the same as the energy computed in the linearized
approximation and is in exact agreement with the expected result. 

\subsection{A ($F$,$D6$) bound state ending on a $D8$-brane}

\noindent
Now let us turn to a little more involved configuration. We will put
the fundamental strings ($F1$) on the $D6$-branes ending on a
$D8$-brane. Similarly to the previous example, the linearized
approximation suggests that the solution in the fullfledged treatment
may be 
\begin{eqnarray}
T&=&{q \over \aprime}\left(x+\cosh\alpha{\pi N\aprime \over r}
\right),\label{tachyon2}\\
B_i&=&{N \over 2}{y_i \over r^3},\label{magnetic}\\
E_i&=&\sinh\alpha {N \over 2}{y_i \over r^3}\label{electric}.
\end{eqnarray}
This could have been obtained by a Lorentz boost in the following way,
though it is not totally clear why it should work in the fullfledged
treatment. Let us think of the fluctuation $\Ttilde$ of the tachyon
$T$ in the linearized approximation as the 9th-component of the gauge
field. Due to the normalization of the tachyon and gauge kinetic
terms, the \lq 9th'-component $\Ttilde$ of the gauge field should be
normalized as ${1 \over q}\Ttilde$.  
Now we start with the configuration (\ref{tachyon}) and (\ref{gauge}),
and perform a Lorentz boost, characterized by an \lq angle' $\alpha$,
in the 9th-direction. Then it gives the solution
(\ref{tachyon2}), (\ref{magnetic}) and (\ref{electric}). In the
linearized approximation we have actually this Lorentz invariance, so
the solutions with nontrivial electric fields should have been
generated in this way\footnote{The zero mode of the fluctuation
  $\tilde{T}$ of the tachyon corresponds to the collective 
  mode transverse to the $D8$-brane, which can indeed be thought 
  of as the dimensional reduction of the 9th-component $A_9$ of 
  the gauge fields.}. 
However there is no apparent Lorentz invariance
of this kind in the fullfledged action, so we are not really entitled
to obtain the above solution. This might be related to the following
subtlety concerning the gauge EOM (the Gauss's law) for the electric
fields. To see it, 
let us first compute the conjugate momentum $\Pi_{A_i}$ of the gauge
field $A_i$. It is given by 
\begin{eqnarray}
\Pi_{A_i}&=&\frac{\delta\calL}{\delta E_i}
=T_{D9}(2\pi\aprime)
e^{-2\pi q^2\left(x+\cosh\alpha{\pi N\aprime \over r}
\right)^2}\sqrt{4\pi^2\aprime}q
\sinh\alpha{\pi N\aprime y_i \over r^3},
\end{eqnarray}
where we used the relation $F[x]=2xF'[x]$ at $x\to\infty$. Thus it
seems at first sight that the Gauss's law (\ref{electricEOM}) 
would imply
\footnote{There is a singularity at $r=0$ where a point charge is 
sitting.} 
\begin{equation}
\del_i\left(\delta\left(x+\cosh\alpha{\pi N\aprime \over r}
\right){y_i \over r^3}\right)=0.
\end{equation}
But apparently this cannot be satisfied. There is, however, a point we
missed, which gives a complete resolution of this problem. Even though
the electric field $E_x$ in the $x$-direction transverse to the
$D8$-brane were not turned on, its conjugate momentum $\Pi_x$ would
not be vanishing due to a particular coupling of the electric field to
the derivative of the tachyon, as we will see below. Now when we
include the electric field $E_x$ in the action (\ref{involved}), the
action is modified to 
\begin{eqnarray}
S&=&-T_{D9}\int dtd^9xe^{-2\pi\aprime T^2}
\sqrt{-
\det\left(\eta_{ab}+2\pi\aprime F_{ab}\right)}\nn\\
&&
F\Biggl[4\pi(\aprime)^2\!\left[
(1+(2\pi\aprime B_i)^2)(\del_xT)^2+(\del_iT)^2
\!-\!(2\pi\aprime\epsilon_{ijk}E_j\del_kT)^2\right.\nn\\
&&\;\;\;\;\;\;
\!+\!(2\pi\aprime B_i\del_iT)^2
-\left\{2\pi\aprime B_i(2\pi\aprime E_i\del_xT
-2\pi\aprime E_x\del_iT)\right\}^2
\nn\\
&&
\;\;\;\;\;\;
\left.
-(2\pi\aprime E_i\del_xT-2\pi\aprime E_x\del_iT)^2\right]/
\left(-\det\left(\eta_{ab}+2\pi\aprime F_{ab}\right)\right)\!
\Biggr],
\label{moreinvolved}
\end{eqnarray}
where $a,b=0,6,7,8,x$, whereas $i,j=6,7,8$, and 
$-\det\left(\eta_{ab}+2\pi\aprime F_{ab}\right)=1+\half (2\pi\aprime
F_{ab})^2-(2\pi\aprime)^4(E_iB_i)^2-(2\pi\aprime)^4 (B_iE_x)^2$. 
Then one can find that the conjugate momentum $\Pi_x$ of the gauge
field $A_x$ is non-vanishing, even when the electric field $E_x$ is
zero: 
\begin{eqnarray}
\Pi_x\!\!\!&=&\!\!\!\frac{\delta\calL}{\delta E_x}
=\!T_{D9}(2\pi\aprime)
e^{-2\pi q^2\left(x+\cosh\alpha{\pi N\aprime \over r}
\right)^2}\!\!\!\sqrt{4\pi^2\aprime}q
\cosh\alpha\sinh\alpha{(\pi N\aprime)^2 \over r^4}.
\end{eqnarray}
Thus the correct EOM for the electric field is
\begin{eqnarray}
\del_i\Pi_{A_i}+\del_x\Pi_x=0,
\end{eqnarray}
which is indeed satisfied by our solution (\ref{tachyon2}),
(\ref{magnetic}) and (\ref{electric}), thanks to a source term
provided by the conjugate momentum $\Pi_x$. 

\parmedskipn
\underline{The energy}

\noindent
Now let us evaluate the energy of this configuration. The Hamiltonian
density $\calH$ is evaluated as 
\begin{eqnarray}
\calH&=&\Pi_{A_i}E_i-\calL\nn\\
&=&T_{D9}
e^{-2\pi q^2\left(x+\cosh\alpha{\pi N\aprime \over r}
\right)^2}\sqrt{4\pi^2\aprime}q
\left(1+\cosh^2\alpha{(\pi N\aprime)^2 \over r^4}\right),
\end{eqnarray}
This gives the energy
\begin{eqnarray}
E&=&T_{D9}\sqrt{2}\pi\sqrt{\aprime}\int d^8y
+T_{D9}2(2\pi^2\aprime)^{3/2}N\cosh\alpha \int d^5ydx\theta(-x).
\end{eqnarray}
Note that $\cosh\alpha=\sqrt{1+\sinh^2\alpha}$ in that $\sinh\alpha$
is proportional to the number $N_{F1}$ of the fundamental strings and
should be quantized as $N_{F1}l/(2\pi\aprime NT_{D6}V_6)$ where $V_6$
and $l$ is the volume of $D6$-branes and the length of the fundamental
strings respectively. Indeed it can be easily checked that this is
precisely the quantization condition of $\Pi_x$. 
Again we find an exact agreement of the energy.

\section{Turning on NS-NS $B$-fields}
\label{turningonB}

\noindent
As a further generalization, let us turn on a constant NS-NS $B$-field
on $D$-branes.  Here we will focus on an interesting phenomenon found
in \cite{Hashimotos} that $D$-branes could be tilted by an effect of
turning on the constant NS-NS $B$-fields. We can convert constant NS-NS 
$B$-fields into constant magnetic fields on $D$-branes. As the 
simplest example, let us consider a $D8$-brane with a constant 
magnetic field, $B_6$, which is equivalent to turning on a constant 
NS-NS $B$-field, $B^{NS}_{78}$, in this case. The tilted $D8$-brane 
will be represented by  
\begin{eqnarray}
T&=&{q \over \sqrt{\aprime}}(x_9-2\pi\aprime B_6x_6),
\label{tiltD8}
\end{eqnarray}
as is obvious. Now we are going to check that this is indeed a
solution of the EOMs. The gauge EOM (\ref{magneticEOM}) is trivially
satisfied. In this case the argument of the functional $F$ is given by 

\begin{eqnarray}
4\pi(\aprime)^2\left((\del_9T)^2+
\frac{(\del_iT)^2+(2\pi\aprime B_i\del_iT)^2}
{1+(2\pi\aprime B_i)^2}\right)
=4\pi\aprime q^2\left(1+(2\pi\aprime B_6)^2\right).
\end{eqnarray}
One can easily find that the tachyon EOM (\ref{tachyonEOM}) boils down
to 
\begin{eqnarray}
&&F[a]=2aF'[a],\\
&&\mbox{with}\qquad a=4\pi\aprime q^2
\left(1+(2\pi\aprime B_6)^2\right).
\end{eqnarray}
Actually this configuration is nothing but the $D6$-$D8$ bound
state. As a check, again let us compute the energy of this
configuration. 
\begin{eqnarray}
E&=&T_{D8}\sqrt{1+(2\pi\aprime B_6)^2}
\int d^7y
d\left(\frac{x_6+2\pi\aprime B_6x_9}
{\sqrt{1+(2\pi\aprime B_6)^2}}\right),
\end{eqnarray}
which is the expected result.

Now let us apply this result to a little more involved case, which is
the tilting\cite{Hashimotos} of the configuration of section
\ref{d6endingd8}, $D6$-branes ending on a $D8$-brane. It is easy to
guess that the solution will be given by 
\begin{eqnarray}
T&=&{q \over \sqrt{\aprime}}\left(x-2\pi\aprime B_6y_6
+{\pi N\aprime \over r}\right),\\
B_i&=&{N \over 2}{y_i \over r^3}+\delta_{i6}B_6.
\end{eqnarray}
One can easily find that the EOMs are satisfied, as can be seen from
the above argument. An important property of the above solution is 
\begin{eqnarray}
\del_iT=-{q \over \sqrt{\aprime}}2\pi\aprime B_i,
\end{eqnarray}
that is again reminiscent of a Bogomol'nyi equation derived in the two
derivative truncation in \cite{KS}. Due to this property the argument
of the functional $F$ reduces to 
\begin{eqnarray}
4\pi\aprime q^2\left(1+(2\pi\aprime B_i)^2\right)
=4\pi\aprime q^2\left(1+(2\pi\aprime B_6)^2
+{(\pi N\aprime)^2 \over r^4}
+(2\pi\aprime)^2 NB_6{y_6 \over r^3}\right).
\end{eqnarray}
Then the square root in the fullfledged action becomes the perfect
square once again, which is indicative of BPS. Hence the energy of
this configuration is evaluated as 
\begin{eqnarray}
E&=&T_{D9}\int d^9x\sqrt{4\pi^2\aprime}qe^{-2\pi q^2
\left(x-2\pi\aprime B_6y_6
+{\pi N\aprime \over r}\right)^2}\nn\\
&&\qquad\qquad\qquad\times
\left(1+(2\pi\aprime B_6)^2
+{(\pi N\aprime)^2 \over r^4}
+(2\pi\aprime)^2 NB_6{y_6 \over r^3}\right),
\end{eqnarray}
perfoming a change of variables, ($\tilde{x}=\frac{x-2\pi\aprime
  B_6y_6}{\sqrt{1+(2\pi\aprime B_6)^2}}$, $\tilde{y}=y_6$), and
adopting a polar coordinate
$(\tilde{y},y_7,y_8)=
r(\cos\phi,\sin\phi\cos\theta,\sin\phi\sin\theta)$, 
we can finally find 
\begin{eqnarray}
E=
T_{D8}\sqrt{1+(2\pi\aprime B_6)^2}\int d^7y
d\left(\frac{y_6+2\pi\aprime B_6x}{\sqrt{1+(2\pi\aprime B_6)^2}}
\right)
+NT_{D6}\int d^5ydx\theta(-x).
\end{eqnarray}
This is exactly the energy of $D6$-branes ending on a tilted 
$D8$-brane.

\parmedskipn
\underline{$D8$-$D6$-$D4$ bound state}

\noindent
As another example, we can also construct a $D8$-$D4$ bound state 
when the $D8$-brane has self-dual noncommutativity on a 4-dimensional 
subspace of its worldvolume. From the Chern-Simons coupling, 
$D4$-branes on $D8$-branes can be described by instantons on the 
$D8$-branes, as is well-known. Usually we have to work on multiple 
$D8$-branes in order to have the instantons on its worldvolume, for the 
$U(1)$ instantons are singular. However when we have self-dual
noncommutativity on the worldvolume, we could have non-singular $U(1)$
instantons as  
in \cite{NS}. So we will turn on a self-dual NS-NS $B$-field, 
$B^+(=B^{NS}_{56}=B^{NS}_{78})$, on a 4-dimensional subspace of a 
single $D8$-brane. Then it is easy to realize that the $D8$-$D4$ bound 
state with $D6$-branes on it can be constructed by
\begin{eqnarray}
T&=&{q \over \sqrt{\aprime}}x_9,\\
A_i&=&B^{NS}_{ij}y_j h(r),
\label{SWinstanton}
\end{eqnarray}
where the indices $i,j$ run from 5 to 8, and the function $h(r)$ of 
$r=\sqrt{y_5^2+y_6^2+y_7^2+y_8^2}$ satisfies the equation 
$2h^2-\left(1+1/(2\pi\aprime B^+)^2\right)h=4N/((B^+)^2r^4)$ 
with $N$ being the number of instantons or $D4$-branes.
The action relevant for this configuration factorizes to 
(\ref{factorize}) with replacing the gauge field strength 
$2\pi\aprime F$ by the invarian field strength 
$\calF=2\pi\aprime (B^{NS}+F)$. Thus the gauge EOMs in this case 
are tantamount to those of the DBI action and the gauge fields 
(\ref{SWinstanton}) given above are nothing but a slight generalization 
 \cite{Tera} of the noncommutative BI-instanton discussed in \cite{SW}.

\section{A junction}
\label{junction}

\noindent
Let us work out another example, a three point junction. We are going
to consider the three point junction of the type
($N_{F1}F$,$-D8$)--($-N_{F1}F$,$0$)--($0$,$D8$). The junction point
could have been pulled away by three types of branes attached their
ends at the junction point. To balance the force at the junction
point, the force vector must be zero and its force balance is simply
determined a la Pythagoras, as indicated in the above notation for the
three point junction. This junction can be realized by ($E_6>0$ for
convenience)\footnote{The junction in this paper is slightly different
  from the one considered in our previous paper\cite{KS}, where we
  used an untilted  ($F$,$D8$) bound state instead of the tilted one
  we employ here.} 
\begin{eqnarray}
\mbox{(I)}\, &T={q \over \sqrt{\aprime}}
(x_9+2\pi\aprime E_6x_6),
\, &E_6=\mbox{const.},\,
(x_6\le 0),\\
\mbox{(II)}\, &T={q \over \sqrt{\aprime}}x_9,\, 
&E_6=0,
\, (x_6> 0).
\end{eqnarray}
We turn on a constant electric flux $E_6$ in the left half ($x_6\le
0$) of the (6,9)-plane by putting negative electric charges along the 
$x_9$-axis and positive charges at one side of the infinity,
$x_6=-\infty$. Our 
junction consists of a ($F$,$D8$) bound state in region (I) and a pure
$D8$-brane in region (II). To balance the force at the junction point
there must be the fundamental strings shooting off from the juction
point, $x_6=x_9=0$, to the negative $x_9$-axis. But we cannot really
see the fundamental strings in our junction solution, while we can
observe the inflow of the fundamental string charges into the
($F$,$D8$) bound state, as we will see below. If we were working on
the $D8$-brane effective theory, there would be no way to see the
fundamental strings which were emanating from far away outside of the
world and abruptly touched down to a point in the world. That was the
case in \cite{DM} in which they worked on the $D1$-brane effectively
theory to consider a three string junction. However we do not only
have the $D8$-brane worldvolume, but also have the bulk of the
spacetime, so we are to be able to see the fundamental strings
manifestly, as is different from the case in \cite{DM}. But we will
leave this problem for future. 

Let us check if the above configuration really satisfies the EOMs. The
solution in region (II) is simply a pure $D8$-brane and thus trivially
a solution of the EOMs, while the one in region (I) contains a subtle
point concerning the EOM for the electric field, as noted in section
\ref{beb}.  

Now starting with a pure $D8$-brane solution in region (II), from the
continuity of the solution at the junction point, we can determine the
coefficient ${q \over \sqrt{\aprime}}$ of $x_9$ in the ($F$,$D8$)
solution in region (I). So the remaning task is to see if the
coefficient ${q \over \sqrt{\aprime}}2\pi\aprime E_6$ of $x_6$ in
region (I) is really consistent with the EOMs, though it is required
physically from the balance of the force at the junction point. One
can readily see that the tachyon EOM (\ref{tachyonEOM}) is indeed
satisfied. To look at the EOM for the electric field (the Gauss's
law), let us first compute the conjugate momentum densities $\Pi_6$
and $\Pi_9$ of the gauge fields $A_6$ and $A_9$ respectively. They are
given by 
\begin{eqnarray}
\Pi_6&=&{\delta\calL \over \delta E_6}
=T_{D8}2\pi\aprime\delta(x_9+2\pi\aprime E_6x_6)
(2\pi\aprime E_6),\\
\Pi_9&=&{\delta\calL \over \delta E_9}
=-T_{D8}2\pi\aprime\delta(x_9+2\pi\aprime E_6x_6)
(2\pi\aprime E_6)^2,
\end{eqnarray}
where again the conjugate momentum density $\Pi_9$ of $A_9$ is not
vanishing, in spite of that the electic field $E_9$ is zero, as noted
in section \ref{beb}. The Gauss's law is thus satisfied: 
\begin{equation}
\del_6\Pi_6+\del_9\Pi_9=0.
\end{equation}
Now we can see the inflow of the fundamental string charges into the
($F$,$D8$) bound state, though we cannot see the fundamental strings
themselves that must be lying along the negative $x_9$-axis. Actually
the charge inflow of the fundamental strings is given by 
\begin{eqnarray}
Q_9=\int d^7y\int_{-\infty}^0dx_6\int_0^{+\infty}dx_9\Pi_9
=-2\pi\aprime (2\pi\aprime E_6)
(T_{D8}V_8),
\end{eqnarray}
where $V_8$ is the volume of the $D8$-brane and this formula gives us
the correct ratio $2\pi\aprime E_6=-(Q_9/2\pi\aprime)/(T_{D8}V_8)$ for
the tensions of the fundamental strings and the $D8$-brane. 

\section{The Emparan-Myers' effect}
\label{emparanmyers}

\noindent
So far we have only dealt with BPS or supersymmetric
configurations. In this section we turn to a nonsupersymmetric
example. We will consider the dielectric $D8$-brane\cite{Emparan,
  Myers} by employing the 
fullfledged action (\ref{involved}). The dielectric $D8$-brane in this
case has the worldvolume of $R^1\times R^6\times S^2$, in which the
attractive force due to the tension of the brane which would shrink
the 2-sphere is cancelled by the flux of the RR 9-form and the
magnetic flux on the $D8$-brane which gives the granular distribution
of $D6$-branes on it. To have the shape of $R^1\times R^6\times S^2$,
the tachyon $T$ may take the form 
\begin{eqnarray}
T={q \over \sqrt{\aprime}}(r-R), \qquad q\to\infty,
\label{tacr}
\end{eqnarray}
where $r$ is a radial coordinate of the (6,7,8)-space, i.e.,
$r=\sqrt{y_6^2+y_7^2+y_8^2}$, while $R$ is the radius of the
2-sphere. Note that, as $q$ goes to infinity, one can see that the
$D8$-brane is localized on the 2-sphere of radius $R$. Also as we have
seen in the linearized approximation, the fluctuation of the tachyon
describes the collective modes transverse to the $D8$-brane. Thus the
dependence on the radius $R$ in the tachyon $T$ can actually be
understood as the collective coordinate in the radial direction. On
the other hand the gauge fields should give the granular distribution
of $D6$-branes, so they may be 
\begin{eqnarray}
B_i=Nc_m{y_i \over r^3}\qquad (F=Nc_md\Omega_2).
\end{eqnarray}
Now the whole action includes the CS term in which we assume only the
RR 9-form $C^{(9)}$ is turned on, as is in an exact analogy with the
Emparan-Myers's effect without tachyon. We give the RR 9-form
$C^{(9)}$ as 
\begin{eqnarray}
C^{(9)}=c\; dt\wedge dV_{R^6}\wedge R^3d\Omega_2,
\end{eqnarray}
where $c$ is a constant parameter.
In this configuration the energy is given by
\begin{eqnarray}
E&=&T_{D9}\int d^6xr^2drd\Omega_2 e^{-2\pi q^2(r-R)^2}
\sqrt{1+\left({2\pi\aprime Nc_m \over r^2}\right)^2}\nn\\
&&\qquad\qquad\times
F\left[4\pi(\aprime)^2\frac{(\del_iT)^2+(2\pi\aprime B_i\del_iT)^2}
{1+\left({2\pi\aprime Nc_m \over r^2}\right)^2}\right]\\
&-&T_{D9}\int e^{-2\pi q^2(r-R)^2}c\; dV_{R^6}\wedge R^3d\Omega_2
\wedge {q \over \sqrt{\aprime}}d(r-R),\nn
\end{eqnarray}
where the last term is the CS term $T_{D9}\int C^{(9)}e^{-2\pi\aprime
  T^2}\wedge dT$. Here one can easily find that the argument of the
functional $F$ simply reduces to 
\begin{eqnarray}
4\pi(\aprime)^2\frac{(\del_iT)^2+(2\pi\aprime B_i\del_iT)^2}
{1+\left({2\pi\aprime Nc_m \over r^2}\right)^2}=4\pi\aprime q^2.
\end{eqnarray}
Thus the energy becomes
\begin{eqnarray}
E=T_{D9}\int d^6x\left[(4\pi)\sqrt{2\pi^2\aprime}
\sqrt{R^4+\left(2\pi\aprime Nc_m\right)^2}
-{1 \over \sqrt{2\aprime}}(4\pi)cR^3\right]
\end{eqnarray}
Indeed this is exactly the same as eq.(87) in \cite{Myers}. Now let us
set $c_m=\half$. Then when the radius $R$ is so small as $R\ll
\sqrt{\pi\aprime N}$, the energy is approximated by 
\begin{eqnarray}
E=T_{D9}2(2\pi^2\aprime)^{3/2}N\int d^6x\left[1+
{1 \over 2(\pi\aprime N)^2}\left(R^4
-NcR^3\right)\right]
\label{dieleE}
\end{eqnarray}
There are two extrema, $R=0$ and $R={3 \over 4}cN$. The latter $R={3
  \over 4}cN$ is the minimum, which indicates the stabilization of the
2-sphere. Note also that the first term in (\ref{dieleE}) is exactly
the energy of $N$ $D6$-branes. 

It is easy to check that the above configuration indeed satisfy the
EOMs. However we would like to remark that the tachyon EOM does not
really give severe restrictions on the form of the tachyon $T$, as
long as the tachyon is of the order of $q$. In fact any tachyon of the
form $T=T(r)$ is allowed. This is somewhat unsual from the viewpoint
of the familiar dynamical systems such as the Abelian-Higgs model. But
this arbitrariness of the tachyon configuration is not as arbitrary as
it stands. Due to the tachyon potential $e^{-2\pi\aprime T^2}$, when
the tachyon is of the order of $q$, only the vicinity of zeros of
$T(r)$ is giving its contribution. Thus we can expand the tachyon
$T(r)$ about zeros $r_0$, which amounts to
$T(r)=T'(r_0)(r-r_0)$. Actually the coefficient $T'(r_0)$ can be
absorbed in the normalization of $q$ and thus it is irrelevant. In
this sense the tachyon configuration (\ref{tacr}) is the \lq unique'
solution of the EOMs. 


\section{A nonabelian example}
\label{nonabelian}

\parmedskipn
\underline{A $D6$-brane suspended between $D8$-branes}

\noindent
Finally we will study the $D$-brane bound states which require
nonabelian extention of previous arguments. As an example we consider
the configuration, a $D6$-brane suspended between two parallel
$D8$-branes which was discussed in the two derivative
truncation\cite{KS}. Let us recall the action (\ref{theaction}) 
\begin{eqnarray}
S&=&-T_{D9}\int dtd^9x\Tr\Biggl(e^{-2\pi\aprime T^2}
\det\sqrt{2\pi}\nn\\
&&\hspace{20mm}\times\left.
\frac{\prod_{r=1/2}^{\infty}\det
\left(\eta_{\mu\nu}+2\pi\aprime F_{\mu\nu}
+4\pi(\aprime)^2{D_{(\mu}TD_{\nu)}T \over r}\right)}
{\prod_{n=1}^{\infty}\det
\left(n\left(\eta_{\mu\nu}+2\pi\aprime F_{\mu\nu}\right)
+4\pi(\aprime)^2D_{(\mu}TD_{\nu)}T\right)}\right).
\end{eqnarray}
As noted before, there is a problem of how to define the ordering of
various $U(N)$ matrices inside the trace. In the two derivative
truncation a certain symmetrized trace prescription appears to be
favorable and we successfully worked out the suspended brane
system. But here we are not trying to resolve this ordering problem,
instead we will simply make an ansatz for the ordering prescription so
that we could deduce a reasonable result. For the relevant
configuration in our concern, we assume the ordering in such a way
that 
\begin{eqnarray}
&&\det\left(\eta_{\mu\nu}+2\pi\aprime F_{\mu\nu}
+4\pi(\aprime)^2{D_{(\mu}TD_{\nu)}T \over n}\right)
=\nn\\
&&
\left(1_{N\times N}+(2\pi\aprime B_i)^2
-(2\pi\aprime E_i)^2-(2\pi\aprime)^2(E_iB_i)^2\right)\\
&&\times\left[1_{N\times N}+{4\pi(\aprime)^2 \over n}
\left((D_9T)^2+\frac{(D_iT)^2
-(2\pi\aprime\epsilon_{ijk}E_jD_kT)^2+(2\pi\aprime B_iD_iT)^2}
{1_{N\times N}+(2\pi\aprime B_i)^2
-(2\pi\aprime E_i)^2-(2\pi\aprime)^2(E_iB_i)^2}\right)\right].\nn
\end{eqnarray}
Then when $E_i=0$, $D_9T={q \over \sqrt{\aprime}}$ and $D_iT=-{q \over 
  \sqrt{\aprime}}2\pi\aprime B_i$, the above determinant reduces to 
\begin{eqnarray}
\left(1_{N\times N}+(2\pi\aprime B_i)^2\right)
\left[1_{N\times N}+{4\pi\aprime q^2 \over n}\left(
1_{N\times N}+(2\pi\aprime B_i)^2\right)\right].
\end{eqnarray}
Further we shall assume the ordering of $U(N)$ matrices so that we
will obtain the action 
\begin{eqnarray}
S&=&-T_{D9}\int dtd^9x\Tr\left[e^{-2\pi\aprime T^2}
\sqrt{4\pi^2\aprime}q
\left(1_{N\times N}+(2\pi\aprime B_i)^2\right)\right],
\end{eqnarray}
which is quite natural, but this is merely an assumption.

Now let us look for a solution which describes a $D6$-brane suspended
between two parallel $D8$-branes. We anticipate that a solution will
be given by the 'tHooft-Polyakov monopole in $SU(2)$ gauge theory
(embedded trivially in $U(2)$). Again it is helpful to invoke the
linearized approximation. We start with a solution of two $D8$-branes,
which takes the form 
\begin{eqnarray}
T={q \over \sqrt{\aprime}}x 1_{2\times 2}.
\end{eqnarray}
The linearized approximation provides us with a $U(2)$ gauge theory
with an adjoint scalar that is the fluctuation $\Ttilde$ of the
tachyon. The potential of the adjoint scalar is absent in this
approximation. So we can obtain the Prasad-Sommerfield limit of the
'tHooft-Polyakov monopole. Thus we conjecture the solution we are
 looking for will be given by\footnote{Here we are employing the
  convention $F_{ij}=\del_iA_j-\del_jA_i-i[A_i,A_j]$.} 
\begin{eqnarray}
T&=&{q \over \sqrt{\aprime}}\left(
x1_{2\times 2}+2\pi \sqrt{\aprime}
{f(r/\sqrt{\aprime}) \over 2r}x_i\sigma_i\right),\\
A_i&=&-i{1 \over \sqrt{\aprime}}{w(r/\sqrt{\aprime})
 \over 2r}x_j\sigma_{ij},
\end{eqnarray}
where the functions $f(r)$ and $w(r)$ take the form
\begin{eqnarray}
f(r)&=&{C \over \tanh(Cr)}-{1 \over r},\\
w(r)&=&{1 \over r}-{C \over \sinh(Cr)}.
\end{eqnarray}
Now let us evaluate the energy of this configuration. To do so, it is
convenient to diagonalize the tachyon $T$, which in particular makes a
$\delta$-function appearing in the energy computation simple and
clarifies the physical picture of this configuration. The tachyon $T$
is diagonalized as 
\begin{eqnarray}
T={q \over \sqrt{\aprime}}\left(
\begin{array}{cc}
x+\pi \sqrt{\aprime}f(r/\sqrt{\aprime}) & 0 \\
0 & x-\pi \sqrt{\aprime}f(r/\sqrt{\aprime})
\end{array}
\right).
\end{eqnarray}
This offers quite a nice physical picture. The function $f(r)$
approaches to zero when $r$ goes to zero, while it monotonically
increases to reach at a constant value $C$ when $r$ is taken to
$+\infty$. Thus the top diagonal element corresponds to a $D8$-brane
located at $x=-\pi \sqrt{\aprime}C$, whereas the bottom one at
$x=+\pi \sqrt{\aprime}C$. Each $D8$-brane has a spike. Two spikes are
shooting off to the center of two $D8$-branes, and eventually
terminate and meet at $x=0$ to compose a tunnel suspended between two
$D8$-branes. 

Now the energy takes the form
\begin{eqnarray}
E&=&T_{D9}\int d^9x\Tr\left[\left(
\begin{array}{cc}
\delta\left(x+\pi \sqrt{\aprime}f(r/\sqrt{\aprime})\right) & 0 \\
0 & \delta\left(x-\pi \sqrt{\aprime}f(r/\sqrt{\aprime})\right)
\end{array}
\right)\right.\nn\\
&&\qquad\qquad\qquad\qquad\qquad\qquad\qquad\times
\sqrt{2\pi^2\aprime}
\left(1_{2\times 2}+(2\pi\aprime B_i)^2\right)\Biggr].
\label{energy868}
\end{eqnarray}
The first term readily gives the energy of two $D8$-branes. So let us
focus on the second term, that is, the energy of a suspended
$D6$-brane. The magnetic energy density $(2\pi\aprime B_i)^2$ can be
computed as 
\begin{eqnarray}
(2\pi\aprime B_i)^2\!\!\!&=&\!\!\!\pi^2\!\left[
{1 \over (r/\sqrt{\aprime})^2}
\left({1 \over (r/\sqrt{\aprime})^2}
-{C^2 \over \sinh^2(Cr/\sqrt{\aprime})}\right)
+G(r/\sqrt{\aprime})\right] 1_{2\times 2},
\end{eqnarray}
where the function $G(r/\sqrt{\aprime})$ is given by
\begin{eqnarray}
G(r/\sqrt{\aprime})&=&4\pi^2{C^2 \over \sinh^2(Cr/\sqrt{\aprime})}
\left[-{1 \over (r/\sqrt{\aprime})^2}
+{C^2 \over \sinh^2(Cr/\sqrt{\aprime})}\right.\nn\\
&&\left.\qquad\qquad\qquad
+2\left({1 \over r/\sqrt{\aprime}}
-{C \over \tanh(Cr/\sqrt{\aprime})}\right)^2\right].
\end{eqnarray}
Thus the energy of a suspended $D6$-brane is
\begin{eqnarray}
E_{D6}&=&4\pi T_{D9}\int d^5y \left[drr^22G(r/\sqrt{\aprime})
+dxd\left(\pi\sqrt{\aprime}f(r/\sqrt{\aprime})\right)
\sqrt{2\pi^2\aprime}\pi\aprime\right.\nn\\
&&\hspace{20mm}\left.\times
\left\{\delta\left(x+\pi\sqrt{\aprime}f(r/\sqrt{\aprime})\right)+
\delta\left(x-\pi\sqrt{\aprime}f(r/\sqrt{\aprime})\right)\right\}
\right].
\end{eqnarray}
Now there is an unwelcome contribution from the function
$G(r/\sqrt{\aprime})$ in the above energy: 
\begin{eqnarray}
I[C]=\int_0^{\infty}dr{C^2 \over \sinh^2(Cr)}\left[
-1+{(Cr)^2 \over \sinh^2(Cr)}
+2\left(1-{Cr \over \tanh(Cr)}\right)^2\right]
\propto C.
\end{eqnarray}
By a simple scaling argument, one can readily find that the integral
$I[C]$ is proportional to $C$, as indicated above. It is also easy to
see that $I[-C]=I[C]$. Thus the unwelcome contribution $I[C]$ is
identically zero, as we anticipated. Thus the energy amounts to 
\begin{eqnarray}
E_{D6}=T_{D6}\int d^5ydx
\left(\theta\left(x+\pi\sqrt{\aprime}C\right)-
\theta\left(x-\pi\sqrt{\aprime}C\right)\right),
\end{eqnarray}
that is exactly the energy of a $D6$-brane suspended between two
$D8$-branes located at $x=-\pi\sqrt{\aprime}C$ and
$x=+\pi\sqrt{\aprime}C$ respectively. 

We have not checked if the above solution really satisfy the EOMs due 
to the ordering problem in the fullfledged treatment, though
remarkably in the two derivative truncation it was successfully
done. The exact agreement of the energy, however, strongly suggests
that it will indeed be a solution of the EOMs. But we will leave it
for future problem. 

\section{Summary and discussions}
\label{discussion}

We extended our previous results in \cite{KS} to the inclusion of
the infinite number of higher derivative corrections for the tachyon
and the gauge fields computed in the BSFT. We find the exact solutions
of the EOM in the fullfledged action (\ref{theaction}) for various BPS 
brane configurations found in \cite{KS}, giving remarkable exact
agreements of the energies and the RR-charges with the expected
superstring results. Among others, we constructed a ($F$,$D6$) bound 
state ending on a $D8$-brane which we could not find in the two
derivative truncation. Indeed it turned out that the existence of this
configuration  
is due to a higher derivative effect in a rather subtle way.
We further discussed the Emparan-Myers' effect
via the tachyon condensation as a nonsupersymmetric example and find
again an exact agreement with the original result of \cite{Myers}
discussed from the effective theory of supersymmetric $D$-branes.  

Although we mostly explored the BPS brane configurations, we have not
really looked at the supersymmetries left unbroken by these
configurations. However it seems quite possible at least to count
fermionic zero modes about our BPS configurations following the
fluctuation analysis of Minahan and Zwiebach\cite{MZ2}. 

We have not considered metamorphoses of the tachyon profile of higher 
codimension $D$-branes whose original linear profiles are given by the
ABS construction $T={q \over \sqrt{\aprime}}\Gamma^ix_i$. As is
apparent from the form of the solution, we need to consider multiple
$D9$-branes in this case and thus we are again facing with the
ordering problem. So the generalization to this case does not seem as
straightforward as one might think. We will leave this problem for
future. 

Putting aside the ordering problem, there is a straightforward 
generalization of the brane configuration studied in section 
\ref{nonabelian}, by simply replacing a suspended $D6$-brane with 
a ($F$,$D6$) bound state. It will be done by adopting the Julia-Zee 
dyon instead of the 'tHooft-Polyakov monopole.
One might think it easy to further extend this configuration to 
$1/4$ BPS configurations\cite{Bergman, hhs, Kawano, hhs2, Kimyeong} 
that are multi-pronged $D1$-branes and ($F$,$D1$) bound states 
suspended between parallel $D3$-branes. In our application it may 
be the simplest case to perform a T-duality in three of four directions 
orthogonal to this configuration, and consider multi-pronged 
$D4$-branes and ($F$,$D4$) bound states suspended between parallel 
$D6$-branes. This is, however, still quite involved, as we need to 
prepare at least two $D9$-branes for each $D6$-brane and further 
to copy a pair of two $D9$-branes as many as the number of $D6$-branes. 
So it is not so easy as one might expect.

Finally we make a comment on the construction of multiple $D8$-branes
out of a single $D9$-brane. As noted in the discussion of the
Emparan-Myers' effect, the tachyon EOM allows us to have rather
arbitrary form of the tachyon $T$. Here we will consider the simplest
example in which only nontrivial field is the tachyon of the form
$T={q \over \sqrt{\aprime}}T(x)$ with $q\to\infty$. It can be easily
checked that arbitrary function $T(x)$ of $x$ satisfies the tachyon
EOM. We will argue that this fact actually makes it possible to
construct multiple $D8$-branes out of a single $D9$-brane.  Let us
assume that $T(x)$ has $N$ zeros $x=x_i$ ($i=1,\cdots,N$). The energy
is readily evaluated as 
\begin{eqnarray}
E&=&T_{D9}\int d^8ydxe^{-2\pi q^2(T(x))^2}
F[4\pi\aprime q^2(T'(x))^2]\nn\\
&=&T_{D9}\sqrt{2\pi^2\aprime}\int d^8y\int_{-\infty}^{+\infty}
\delta(T(x))|T'(x)|dx\nn\\
&=&\sum_{i=1}^N T_{D9}\sqrt{2\pi^2\aprime}
\int d^8y\int_{-\infty}^{+\infty}\delta(x-x_i)dx\nn\\
&=&NT_{D8}\int d^8y,
\end{eqnarray} 
that is exactly the energy of $N$ $D8$-branes. From the CS coupling,
however, it is easy to see that the sign of $T'(x_i)$ corresponds to
that of the $D8$-brane charge, so the above configuration is either
$N/2$ $D8$-$N/2$ anti$D8$ pairs or $(N-1)/2$ $D8$-$(N-1)/2$ anti$D8$
pairs with leaving a single $D8$ (or anti$D8$) unpaired, when $T(x)$
is a polynomial function. More interesting solution is of
nonpolynomial types, for instance, $T(x)=\tan x$. Then we could have
multiple $D8$-branes without having multiple numbers of $D9$-branes.

\section*{Acknowledgements}
S.\ H.\ is indebted to N.\ Sasakura for discussions. K.\ H.\ 
and S.\ H.\ were supported in part by the Japan Society for the
Promotion of Science.  This research was supported in part by the
National Science Foundation under Grant No.\ PHY99-07949.


\end{document}